\documentclass[pra,twocolumn]{revtex4}
\usepackage{graphics}
\usepackage{amsmath,amssymb}
\newcommand{\abs}[1]{\lvert#1\rvert}
\newcommand{\eps}{\epsilon}

\begin{document}
\title{St\"uckelberg--Interferometry with ultra-cold atoms}
\author{Patrick Pl\"otz and Sandro Wimberger
}                     

\affiliation{Institut f\"ur Theoretische Physik, Universit\"at Heidelberg, Philosophenweg 19, 69120 Heidelberg}
\date{\today}
%
\begin{abstract}
We show that and how ultra-cold atoms in an accelerated two-band lattice are a controlled realization of  Landau--Zener--St\"uckelberg interferometry. 
\end{abstract}
%

\maketitle

\section{Introduction}\label{intro}
Two-level systems subject to a strong periodic driving appear in a large variety of quantum mechanical systems and their study has a long history. They are naturally used in atomic physics when an atom is coupled to a laser-field~\cite{grifoni}, in solid state physics when describing superconducting qubits~\cite{squbit} or as effective models~\cite{Longhi}. One consequence of the strong driving is the possibility for the two-level system to undergo a sequence of transitions. Each transition can be seen as an effective beam splitter and the coherent passage through several transitions leads to an accumulation of phases and interference effects known as St\"uckelberg oscillations~(see \cite{shevchenko} and references therein). 

Recently, St\"uckelberg oscillations have been observed experimentally for ultra-cold atoms in accelerated optical lattices~\cite{arimondo,salger}. This opens the route for very detailed studies of St\"uckelberg interferometry with cold gases. The high degree of control in these systems~\cite{OberthalerReview,BlochZwergerReview} allows to explore the strong sensitivity of the phase between interband transitions on the band structure. The main goal of the present paper is to establish explicitly the connection between St\"uckelberg interferometry and interband transitions in optical lattices. Additionally, we provide simple analytical formulae for the interband dynamics and compute interference patterns, i.e. contour plots of transition probabilities, for realizations with a single optical lattice as experimentally used in~\cite{arimondo} and superlattices as in~\cite{salger}. \\
The outline is as follows. In section~\ref{sec:analogy} we are going to obtain the Landau--Zener--St\"uckelberg (LZS) Hamiltonian from a two-band model for ultra-cold atoms in accelerated optical lattices and study the dynamics of the interband transitions using a systematic expansion in section~\ref{sec:magnus}. We will then compute the transition probabilities using degenerate perturbation theory and compare the results to numerical simulations. This will be followed by predictions for interference patterns in realizations with a single optical lattice and superlattices. We will close with a short summary.

\section{Cold atom realization of the Landau--Zener--St\"uckelberg Hamiltonian}\label{sec:analogy} 
A quantum mechanical two-level system with energy bias $\varepsilon_0$ under strong periodic driving with amplitude $A$ and frequency $\omega$ is modeled by the Landau--Zener--St\"uckelberg Hamiltonian~\cite{shevchenko} 
\begin{equation}\label{eq:lzs}
	 \mathcal{H}_{\text{LZS}} = -\frac{1}{2}\left( 
	 \begin{array}{cc} \varepsilon_0 + A\sin \omega t  & \Delta_T \\
	 \Delta_T & -\varepsilon_0 - A\sin \omega t 	 \end{array} \right).
\end{equation}	
Here, $\Delta_T$ denotes the tunneling amplitude between the two levels. In the present section we are going to show explicitly how this Hamiltonian can be realized with ultra-cold atoms in accelerated optical lattices. 

Using ultra-cold atoms in optical lattices it is possible to create tilted non-interacting two-band systems~\cite{salger}. 
We make all parameters of the Hamiltonian dimensionless by measuring them in units of recoil energies $E_{\rm rec}\equiv \hbar^2k_L^2/(2m)$, where $k_L$ is the wave vector of the laser creating the optical lattice and $m$ the mass of the atoms (we set $\hbar = 1$). The dimensionless force $F$ is obtained by multiplication of the physical force with the lattice constant and dividing by $E_{\rm rec}$. The appropriate dimensionless two-band Hamiltonian obtained from an expansion in Wannier functions of the full problem~\cite{TMW1,andrea} reads
\begin{align}\label{eq:fullHamiltonian}
& \mathcal H_{\text{2B}} = \sum_{l\in \mathbb Z} \Big[\big(lF -\tfrac{\Delta}{2}\big) a_l^{\dagger}a_l^{} - \frac{J_a}{2}(a_{l+1}^{\dagger}a_l^{} + \mathrm{h.c.}) \\ 
& +\big(lF+\tfrac{\Delta}{2} \big) b_l^{\dagger}b_l^{} - \frac{J_b}{2}(b_{l+1}^{\dagger}b_l^{} + \mathrm{h.c.}) + FC_0(b_l^{\dagger}a_l +  \mathrm{h.c.}) \notag \Big]. 
\end{align}
The operator $a_l$ ($a_l^{\dag}$) annihilates (creates) a particle at site $l$ and $b_l$ ($b_l^{\dag}$) in the upper band. The bands are separated by a bandgap $\Delta$ and the whole lattice is tilted by on-site energies $lF$. The hopping amplitudes between neighbouring sites in band $a, b$ are denoted by $J_a>0$ and $J_b<0$. The single-particle coupling of the bands is proportional to the external Stark force $F$ via $C_0 F$ with a coupling constant $C_0$ depending on the optical lattice but usually of order $C_0\approx -0.2$~\cite{TMW1,andrea}. We take the external force $F$ as a free parameter. 
The parameters of the Hamiltonian are directly computed from Wannier functions $w_l^{a,b}(x)$, which are maximally localized states centered around the $l$-th lattice well. They can be computed for realizations with a single optical lattice~\cite{andrea} $V(x) = V_0 \cos(x)$ or with a superlattice $V(x) = V_1\cos(x)+V_2\cos(2x+\phi)$ with a possible additional phase $\phi$ between the lattices~\cite{diss,carlos}. The parameters are then given by
\begin{subequations}\label{eq:parameters}
\begin{align}
	J_{a,b} & = \int w^{a,b}(x) V(x) w^{a,b}(x) \mathrm d x \\
	C_0 & = \int w^{a}(x) \cdot x\cdot  w^{b}(x) \mathrm d x .
\end{align}
\end{subequations}
Here, $ w_l^{a,b}(x) $ denote the Wannier functions for the lowest ($ a $) and first excited band ($ b $) that are computed from the Bloch functions to the potential $ V(x)$. The band gap $ \Delta $ is the difference between the average energy of these two lowest Bloch bands. Inter-particle interactions can be neglected under specific experimental conditions~\cite{arimondo} and the case of weak interactions can be treated perturbatively, extending the discussion in~\cite{collapse,spinmodel}.

To obtain a driven time-dependent two-level system similar to eq.~(\ref{eq:lzs}), we change to the interaction-picture with respect to the external force~\cite{FB}. This removes the tilt $\sum_l lFa_l^{\dagger}a_l^{}$ and replaces $a_{l+1}^{\dagger}a_l^{}\rightarrow {e}^{i Ft}a_{l+1}^{\dagger}a_l^{}$ (and likewise for $b_{l+1}^{\dagger}b_l^{}$). The Hamiltonian is then time-dependent with a periodicity of $T_B\equiv 2\pi/F$. Introducing Fourier components 
$a(k) = \sum_l e^{i lk}a_l$ and $b(k) = \sum_l e^{i lk}b_l,$ 
we obtain the following periodic two-level Hamiltonian~\cite{Zhao,noninter}
\begin{equation}\label{eq:Hkspace}
	 \mathcal{H} = \begin{pmatrix}
	 -\frac{\Delta}{2} - J_a \cos(k+Ft) & C_0F   \\
			C_0 F &  \frac{\Delta}{2} - J_b \cos(k+Ft) 
	 \end{pmatrix}.
\end{equation}	
The finite distance between the two levels, a sinusoidal driving and a constant coupling as in the LZS Hamiltonian, eq.~(\ref{eq:lzs}), are already present now. To make the connection completely transparent, we add a periodic shift of the energy zero. The final result is then
\begin{multline}\label{eq:coldatomlzs}
 \mathcal{H} =  -\frac{1}{2}\begin{pmatrix}
	 \Delta +2 J \cos(k+Ft) & -2C_0F   \\ -2 C_0 F &  -\Delta - 2 J \cos(k+Ft)  \end{pmatrix}  \\ 
	 +  (J_a+J_b) \cos(k+Ft) \begin{pmatrix} \;1\; & \;0\; \\ 0 &1 \end{pmatrix} ,
\end{multline}	
where $J=J_a-J_b$. Shifting the time zero as $t\rightarrow t - k/F - \pi/2F$, we arrive exactly at the form of eq.~(\ref{eq:lzs}). The first part of the Hamiltonian eq.~(\ref{eq:coldatomlzs}) is the ultra-cold atom realization of the Landau--Zener--St\"uckelberg Hamiltonian and the second part reflects a time-dependent shift of the zero energy point, which does not concern interference effects between different phases. To simulate the LZS Hamiltonian and to perform interferometry one can use the control offered by ultra-cold atom systems in accelerated optical lattices. The role of the energy bias is taken by the average band gap between the two Bloch bands, the driving amplitude is realized as the difference in hopping strengths and the driving frequency is the Bloch frequency $\omega_B = F$ in our units. Finally, the tunneling amplitude in the ultra-cold atom realization is proportional to the external Stark Force. This correspondence between the two realizations is summarized in table~\ref{tab:corr}.
\begin{table}[t]\centering
\begin{tabular}{cc}
\hline tilted optical lattice & Landau--Zener--St\"uckelberg \\\hline 
\hline  band gap $ \Delta $ & level splitting $ \varepsilon_0 $ \\ 
	Bloch freq. $ \omega_B=F $ & driving freq. $ \omega $ \\ 
	band coupling $2 C_0 F$ & level coupling $ -\Delta_T/2 $ \\ 
	hopping $ J = J_a-J_b$ & field strength $ A $ \\ 
\hline 
\end{tabular} \caption{Analogy between tilted optical lattices and the Landau--Zener--St\"uckelberg Hamiltonian. Note that the band coupling and the driving (or Bloch-) frequency are not independent in tilted optical lattices.}\label{tab:corr}
\end{table}
Please note that not all parameters in the cold atom realization can be varied independently since the driving frequency $\omega_B = F$ and the interband coupling $C_0 F$ both depend on the external force.
 \begin{figure}
	\centering
	\resizebox{0.475\textwidth}{!}{\includegraphics{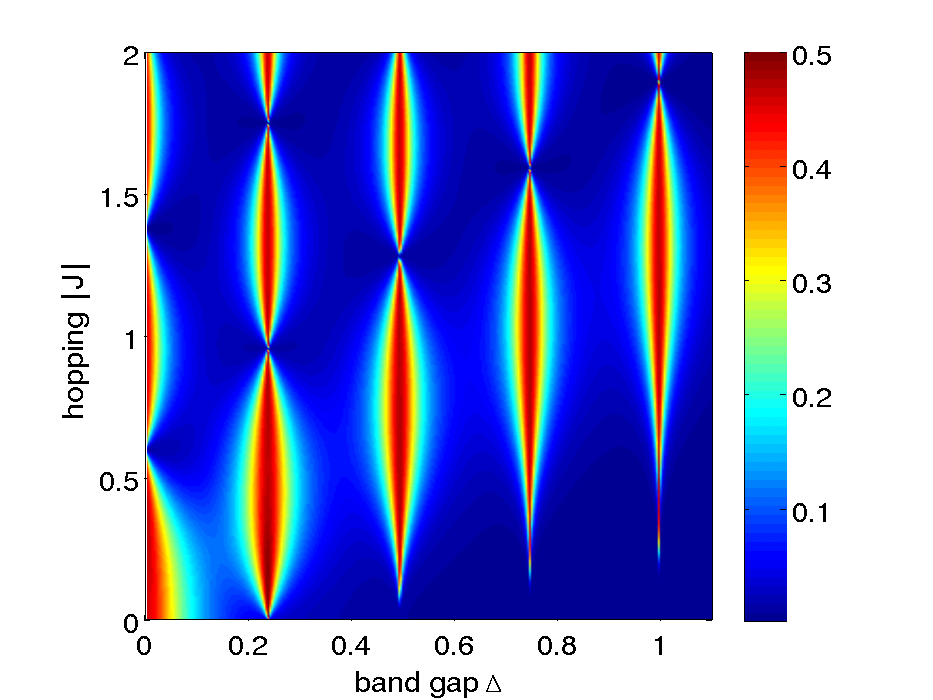}}
	\caption{(Color online) Landau--Zener--St\"uckelberg interferometry. Shown is the long-time average of the occupation of the upper band as a function of the band gap $ \Delta $ (corresponding to the level separation in atomic systems) and the hopping difference between the bands $ J=J_a-J_b $ (corresponding to the driving amplitude). Parameters: $C_0 = -0.15$ and $F=1.0$.}
	\label{fig:inter}       
\end{figure}
Figure~\ref{fig:inter} presents a typical LZS interferometric pattern (see~\cite{shevchenko} for many examples and a summary of different experimental results). Shown are the transition probabilities, i.e. the long-time average ($t\ggg T_B$) of the occupation of the upper band, when varying the 'driving strength' $J$ and the 'level splitting' $\Delta$ according to eq.~(\ref{eq:occtot}) to be derived below. To see how these interference patterns arise in systems of cold atoms, we will derive analytical expressions for the interband dynamics.

	\section{St\"uckelberg Interferometry with ultra-cold atoms}\label{sec:magnus}
\subsection{Dynamics of the interband transitions}
To study the dynamics of the interband transition we have to solve the time-dependent Schr\"odinger equation. However, it can be shown to be equivalent to the Hill equation and exact analytical solutions in closed form are not possible. To obtain approximate solutions in a systematic fashion, we will use the Magnus expansion~\cite{MagnusReview}. It is useful apply the following transformation in order to obtain a purely off-diagonal Schr\"odinger equation~\cite{Zhao,noninter}  \begin{subequations}
\begin{align}
	\tilde a(k,t) & =  a(k,t) \mathrm e^{-i\Delta\cdot t/2 - i J_a \int_0^t \cos(k + Ft')\mathrm{d} t'}\\
	\tilde b(k,t) & =  b(k,t)\mathrm e^{+i\Delta\cdot t/2 - i J_b \int_0^t \cos(k + Ft')\mathrm d t'}. 
\end{align}
\end{subequations}
This removes the diagonal terms and we obtain an equivalent Schr\"odinger equation with a transformed Hamiltonian $\tilde{\mathcal H} (t)$ for the transformed amplitudes 
\begin{equation}\label{eq:Schroedinger}
	i\frac{\partial}{\partial t} \binom{\tilde a(k,t)}{\tilde b(k,t)} = \begin{pmatrix} 0 & C_0Fe^{-i\phi(k,t)} \\ 
		C_0Fe^{i\phi(k,t)} & 0 \end{pmatrix}\binom{\tilde a(k,t)}{\tilde b(k,t)},
\end{equation}
where $\phi(k,t) = \Delta \cdot t - (J/F)[\sin(k+Ft) - \sin(k)]$ is the phase between the two Bloch bands and $J = (J_a-J_b)$. Where the pure existence of two energy bands allows phenomena like Rabi oscillations, it is the non-trivial phase difference for the time-evolution in both bands, related to $J \neq 0$, that gives rise to St\"uckelberg oscillations and the complex interference phenomena. That is, the difference in the curvature $J = J_a-J_b \neq 0$ allows to collect different phases during the time evolution in the upper or lower band and excludes exact analytical solutions of the Schr\"odinger equation in closed form~\cite{grifoni,shevchenko,noninter}. 

The idea of the Magnus expansion is to express the time evolution operator as the exponential of an infinite series $U(t)=\exp \left[ \sum_{n=1}^{\infty}\Omega_{n}(t) \right], $
with each term containing an increasing number ($n-1$) of nested different time commutators~\cite{MagnusReview}. All orders of the Magnus expansion could be summed in a recent work for a new derivation of the transition probability in the Landau--Zener problem~\cite{rojo}. The first two terms of this series read explicitly (valid in both Schr\"odinger and interaction picture~\cite{MagnusPechukas})
\begin{subequations}
\begin{align}
	\Omega_1(t) &  =  -i \int_0^t \tilde{\mathcal H}(t_1)~\text{d}t_1 \\
	\Omega_2(t) &  =  \frac{1}{2}\int_0^t \!\mathrm d t_1 \int_0^{t_1}\! \mathrm d t_2 
	\; [\tilde{\mathcal H}(t_1),\tilde{\mathcal H}(t_2)] 
\end{align}
\end{subequations}
To compute the terms in the Magnus expansion for our time-dependent problem, eq.~(\ref{eq:Schroedinger}), explicitly, we have to integrate the Hamiltonian and commutators of it over time. Without loss of generality, we restrict our discussion to $k=0$, which can always be achieved by shifting the time zero. We will need the integral
\begin{equation}
	\chi(t) \equiv \int_0^{t} e^{i \phi(t')}\mathrm d t' = \int_0^t e^{i\left(\Delta\cdot t' - (J/F)\sin Ft'\right)} \mathrm d t'.
\end{equation}
Using the the generating function of the Bessel function $\mathrm{exp}[u (\tau-1/\tau)] = \sum_{n\in \mathbb Z} J_n (2 u) \tau^n$
for $\tau = e^{-i Ft}$, we can write the $\sin$ in the exponent as a sum over Bessel functions
$\chi(t) = \sum_{n\in \mathbb Z} J_n (J/F) \int_0^{t}\! \mathrm d t' e^{i (\Delta -nF)t'}.$
After integration and minor manipulations, we obtain a closed expression with the explicit time dependence for the highly oscillatory function
\begin{equation}\label{eq:chi}
	\chi^{}(t) = 2 \sum_{n\in\mathbb Z} J_n(J/F)\; e^{i \omega_n t/2} \; \frac{\sin\, (\omega_n t/2)}{\omega_n}
\end{equation}
with $\omega_n = \Delta - nF$. With the explicit expression for the integral defining $\chi(t)$, the first term in the Magnus expansion is given by $\Omega_1(t)  =  -i C_0F \bigl(\begin{smallmatrix} 0 & \chi^{*} \\ \chi & 0 \end{smallmatrix}\bigr)$ and the time evolution operator in first order reads correspondingly~\cite{diss}
\begin{equation}
	U_1(t) =
	\begin{pmatrix} \cos (C_0F |\chi|) & -ie^{i \arg \chi} \sin (C_0F |\chi|) \\ 	-ie^{-i \arg \chi} \sin (C_0F |\chi|) & \cos (C_0F |\chi|) \end{pmatrix} .
\end{equation}
This is the result in first order and the occupation of the upper band $P_b(t) \equiv \sum_k \abs{b(k,t)}^2$ is given by 
\begin{equation}\label{eq:mag1}
 P_b(t)= \sin^2\bigg[2C_0F\Big| \sum_n J_n(J/F) 
	e^{i \omega_n t/2} \; \frac{\sin\, (\omega_n t/2)}{\omega_n} \Big|\bigg].
\end{equation}
This result captures resonant and non-resonant contributions of the interferometry in a single and explicit  formula. Eq.~(\ref{eq:mag1}) can be understood by treating the denominator zeros $\omega_m=0\Leftrightarrow\Delta = mF$ contained in $\chi(t)$ separately using $\lim_{x\rightarrow0}\frac{\sin xt}{x} = t$. This condition $\Delta\approx mF$ corresponds to a resonant interband coupling and $\chi(t)$ can be decomposed as
\begin{equation*}
	\chi^{}(t) = J_m(J/F)\, t + 2 \sum_{n\neq m} J_n(J/F)\; e^{i \omega_nt/2} \; \frac{\sin\, (\omega_n t/2)}{\omega_n} .
\end{equation*}
For large times the first term will be dominating and the overall short-time averaged occupation of the upper band shows large sinusoidal oscillations with unit amplitude $\mathcal N_b^{\text{res}}(t) = \sin^{2} [VJ_m(J/F)\, t ].$
The other high-frequency and non-resonant terms lead to small amplitude oscillations on top of this overall resonant interband oscillations. 

For the second order contribution we need the commutator (where $\sigma_z$ is the diagonal Pauli matrix)
\begin{equation}
	[ \tilde{\mathcal H} (t_1), \tilde{\mathcal H} (t_2)] =  2i C_0^2F^2  \sigma_z \sin[\phi(t_2)-\phi(t_1)]
\end{equation}
and integrate over it in time. One obtains $\Omega_2(t)  = i C_0^2F^2 \sigma_z \psi(t)$ where the required integral $\psi(t) \equiv \int_0^t \!\mathrm d t_1 \int_0^{t_1}\! \mathrm d t_2 \sin[\phi(t_2)-\phi(t_1)]$ can be computed by applying the same expansion as above
\begin{equation}\begin{split}
	\psi(t) = & \sum_{n\in\mathbb Z} \frac{J_n(J/F)}{\omega_n}
		\Big\{   \sum_{m\in\mathbb Z}J_m(J/F)\times \\ 
	& \Big( \frac{\sin^2\omega_mt/2}{\omega_m} -  \frac{\sin^2[(m+n)Ft/2]}{(m+n)F}\Big) \Big\}.
\end{split}
\end{equation}
The time evolution operator can again be given exactly~\cite{diss} and one finds for the occupation of the upper band
\begin{equation}\label{eq:mag2}
	P_b(t) = \frac{\abs{\chi^{}}^2}{\abs{\chi^{*}\chi^{}+C_0^2F^2\psi^{2}}} 
		\sin^2 \Big(2C_0F \sqrt{\chi^{*}\chi^{}+C_0^2F^2\psi^{2}}\,\Big),
\end{equation}
where we suppressed the time-dependence of the functions $\chi(t)$ and $\psi(t)$. The eqs.~(\ref{eq:mag1}) and~(\ref{eq:mag2}) are the central results of this section. They provide good approximations to the full interband dynamics in an explicit expression. They furthermore contain the resonant as well as non-resonant contributions to the interband dynamics.

\subsection{Average occupation of bands}
In order to determine the average occupation of the bands it is more useful to go back to the original two-band Hamiltonian eq.~(\ref{eq:fullHamiltonian}) instead of calculating the long-time average over the expressions eq.~(\ref{eq:mag1}) and eq.~(\ref{eq:mag2}). Introducing the Wannier--Stark states~\cite{WSreview} $\alpha_n  = \sum_{l\in\mathrm Z} J_{l-n}(J_a/F) a_l$ and $\beta_n  = \sum_{l\in\mathrm Z} J_{l-n}(J_b/F) b_l$ into eq.~(\ref{eq:fullHamiltonian}), one obtains~\cite{collapse}
\begin{multline}\label{eq:transformedH}
\mathcal H_{\text{2B}} = \sum_{l\in\mathrm Z}\Big[ \big(lF -\tfrac{\Delta}{2}\big) \alpha_l^{\dagger}\!\alpha_l
 +\big(lF +\tfrac{\Delta}{2}\big)\beta_l^{\dagger}\!\beta_l  \\
	+ \sum_{m\in \mathrm Z} C_0F  J_{m}(J/F) ( \alpha_l^{\dagger}\!\beta_{l-m}  + \rm{h.c.} ) \Big],
\end{multline}
where $J = J_a - J_b$ as above. This expression contains all relevant processes coupling the two bands as direct couplings weighted by Bessel functions $J_m(J/F)$. For \mbox{$J/F\lesssim1$} the dominant contribution is the on-site coupling between the bands proportional to $J_0(J/F)$. Keeping only this dominant contribution, the Hamiltonian eq.~(\ref{eq:transformedH}) is a sum of independent two-level systems and can easily be diagonalized. The resulting occupation of the upper band contains the dominant part of the non-resonant oscillations already contained in eq.~(\ref{eq:mag1}) and reads
\begin{equation}
 	P_b(t) = \frac{4V_0^2}{\Delta^2+4V_0^2}\sin^2\Big(\sqrt{\Delta^2+4V_0^2}\cdot t/2 \Big).
\end{equation}
where $V_0 = C_0FJ_0(J/F)$. This means that the non-resonant contribution to the averaged occupation of the upper band is given by
\begin{equation}\label{eq:meanocc}
\overline{P_b} = \frac{1/2}{1+ \big[\frac{\Delta/F}{2C_0J_0(J/F)}\big]^2 }.
\end{equation}

However, in addition to this onsite coupling, the Hamiltonian eq.~(\ref{eq:transformedH}) allows a direct coupling of more remote sites. This becomes particularly important whenever a site from the lower band and a site from the upper band are energetically degenerate. The corresponding resonance  condition for two levels being separated by $m$ sites is $\Delta \approx mF$, $m\in\mathbb N$. We therefore apply degenerate perturbation theory~\cite{ShirleyThesis} to the Hamiltonian eq.~(\ref{eq:transformedH}) and obtain for a resonance of order $m$ in second order the following effective two-level system~\cite{diss,twolevelsystem} 
\begin{equation}\label{eq:pt}
\begin{pmatrix} 
\eps_{l-m}^{+} +  {\displaystyle \sum_{i\neq l}\frac{|V_{l-m-i}|^2}{\eps_{l-m}^{+}-\eps_l^-}} & V_{-m} \\
	 V_{-m}  & \eps_{l}^{-}+  {\displaystyle \sum_{i\neq l-m}\frac{|V_{l-i}|^2}{\eps_{l}^{-}-\eps_i^+}} 
\end{pmatrix},
\end{equation}
with $\eps_l^{\pm} = lF\pm \Delta/2$ and $V_{l} = C_0FJ_{l}(J/F)$. This and higher orders  allow the computation of various observables with high degree of precision~\cite{twolevelsystem}. For example, the resonance condition $\Delta = mF$ experiences a slight Stark shift and the corresponding condition in second order is given by
\begin{equation}\label{eq:res}
 	\Delta = mF - 2C_0^2F^2\sum_{i\neq m}\frac{J_i^{2}(J/F)}{\Delta-iF}.
\end{equation}
Unlike the usual LZS problem, the coupling between the bands $C_0F$ and the driving frequency $\omega_B = F$ are not independent for atoms in optical lattices. This makes eq.~(\ref{eq:res}) nonlinear and difficult to solve. However, it can be solved either numerically or by iteration. The uncorrected resonance position for a single optical lattice with $V_0 = 4$ and a resonance of order 2 is given by $F_2 = \Delta/2=2.195$. Solving eq.~(\ref{eq:res}) numerically gives $F_2^{\text{PT}} = 2.22067$ which has a relative error of order $10^{-5}$ when compared to the maximum of the resonance $F=2.22070$ from numerical simulations of the full problem eq.~(\ref{eq:Hkspace}). In the same way, very high precision can be achieved by higher orders perturbation theory.

\begin{figure}[t]
	\centering
	\resizebox{0.475\textwidth}{!}{\includegraphics{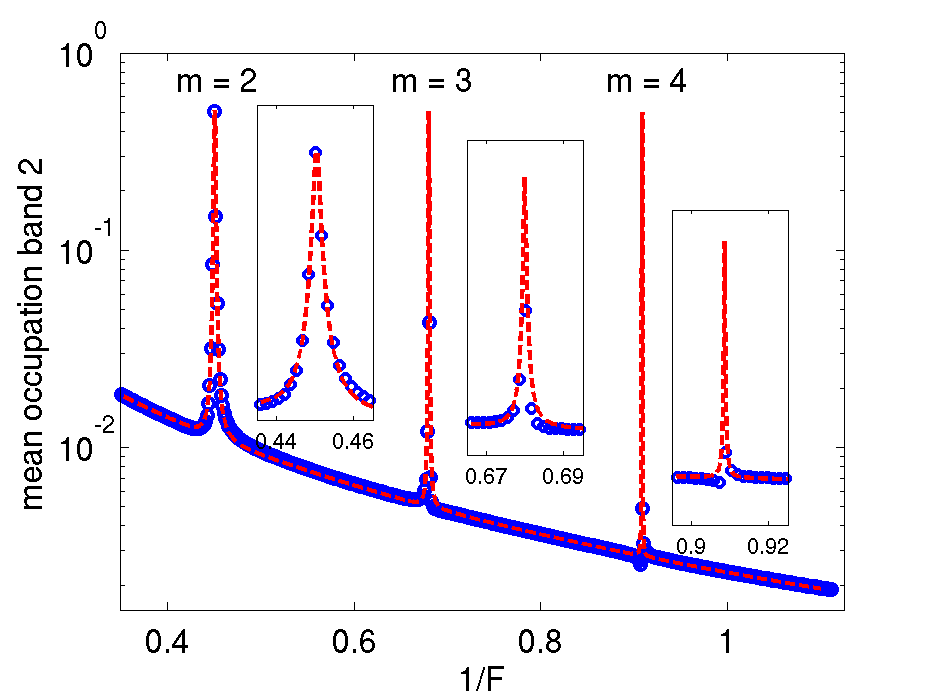}}
	\caption{(Color online) Longtime average of the occupation of the upper band from numerical simulations of the full problem (blue circles) and the theoretical prediction eq.~(\ref{eq:occtot}) (dashed line). The insets show the resonances $m=2,3,4$ on a linear scale. Parameters for a single optical lattice of depth $V_0 = 4$: $C_0 = -0.14$, $\Delta = 4.39$, and $J= -0.682$.}
	\label{fig:meanocc}       
\end{figure}

The resulting resonant contribution to the average occupation of the upper band has a Lorentzian shape~\cite{shevchenko,noninter} 
and the total transition probability is given by the non-resonant interband coupling and the different resonant contributions
\begin{equation}\label{eq:occtot}
 	\overline{P_b} = \frac{1}{2}\frac{4V_0^2}{\Delta^2+4V_0^2} 
		+\frac{1}{2} \sum_m \frac{4(V_m/F\Delta)^2}{(1/F-1/F_m)^2+4(V_m/F\Delta)^2}.
\end{equation}
The first term describes the direct force-induced coupling between the bands and is usually much smaller than unity. However, for $F\gg \Delta$ its contribution becomes important and grows as $F^{-2}$. The second part are the resonant contributions from different orders of resonance. 
Let us compare this result eq.~(\ref{eq:occtot}) to numerical simulations of the full Hamiltonian eq.~(\ref{eq:fullHamiltonian}). We change to the interaction picture with respect to the external force and impose periodic boundary conditions~\cite{FB,andrea}. We take an initial state $|\psi_0\rangle$ which occupies only the lower band, and evolve it in time according to the Schr\"odinger equation $i\partial_t|\psi(t)\rangle = \mathcal H(t)|\psi(t)\rangle$. We compute the total occupation of the upper band $P_b(t)=\langle\psi(t)|b_l^{\dagger}b_l^{}| \psi(t)\rangle$ and take the long time average. The result for a single optical lattice realization with $V_0=4$ is shown in fig.~\ref{fig:meanocc} together with our result eq.~(\ref{eq:occtot}). 
We observe very good agreement even on the logarithmic scale shown in the figure. The non-resonant interband coupling and the resonant contribution as well as the resonance positions are accurately reproduced. Only the asymmetry of the resonance peaks (in particular of the high order $m=4$ resonance at $1/F\approx 0.9$) is not captured by our present analysis since the effective model of eq.~(\ref{eq:pt}) should be extended to describe well such more complex peak profiles.

\subsection{Interferometry with optical lattices}
\begin{figure*}
	\centering
	\resizebox{0.45\textwidth}{!}{\includegraphics{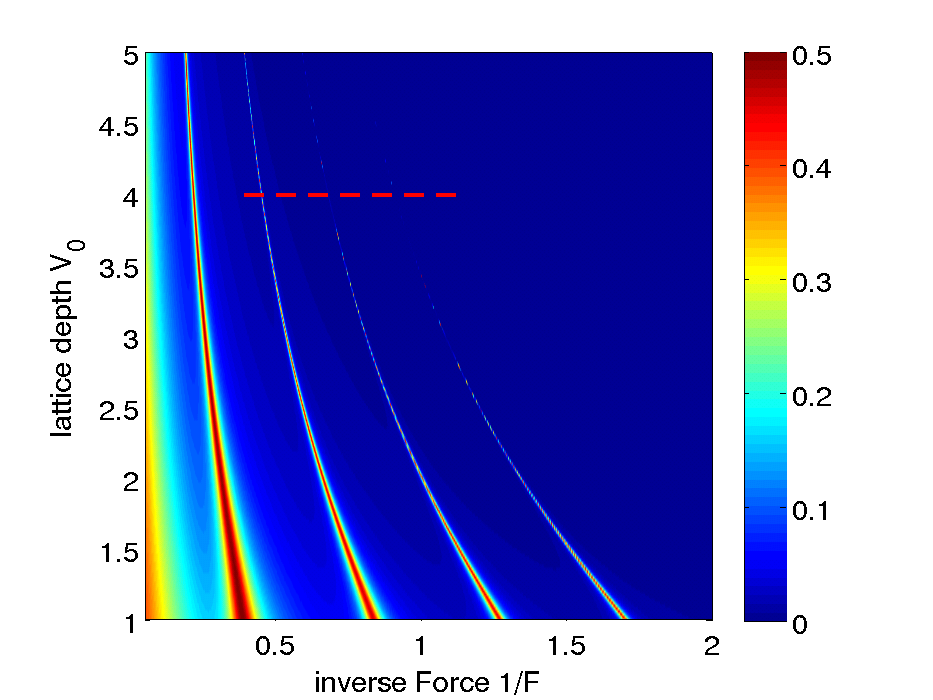}}\hspace{0.05\textwidth}
	\resizebox{0.45\textwidth}{!}{\includegraphics{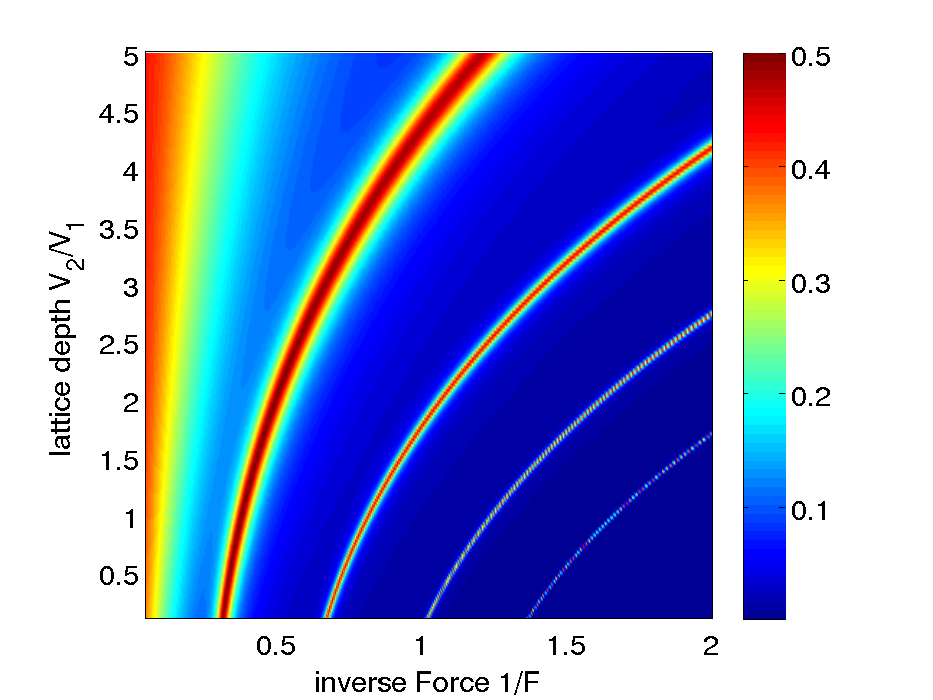}}
	\caption{(Color online) Driving interferometry, i.e. average occupation as function of the upper band as a function of the lattice depth and the external force. \emph{Left:} Long-time average of the occupation of the upper band according to eq.~(\ref{eq:occtot}) for a realization with a single optical lattice with depth $V(x)=V_0\cos(x)$. The parameters $J=J_a-J_b$ and $C_0$ depend on the lattice depth and are determined from numerical computation of the Wannier functions and eq.~(\ref{eq:parameters}). The dashed line marks the values shown in fig.~\ref{fig:meanocc}. \emph{Right:} The same as in the left panel but for a realization with a superlattice $V(x)=V_1\cos(x)+V_2\cos(2x+\phi)$ for $V_1 = 2$, $\phi=\pi$, and varying $V_2$.}
	\label{fig:singlelattice}       
\end{figure*}

We have established the general possibility to use ultra-cold atoms for St\"uckelberg interferometry and have given analytical results for the probability of interband transitions in the previous paragraphs. As already mentioned, the important parameters hopping $J$ and band gap $\Delta$ to build the Landau--Zener--St\"uckelberg Hamiltonian, eq.~(\ref{eq:lzs}), can, however, not be  controlled independently in optical lattice systems. We therefore show interference patterns similar to figure~\ref{fig:inter} but with the experimentally accessible parameters varied. That is, we vary the depth of the optical (super-)lattice (and possible the phase $\phi$) and compute the Wannier functions for each value of the lattice parameters. From these we obtain the relevant parameters $J$, $\Delta$, and $C_0$ according to eq.~(\ref{eq:parameters}). With them we obtain the transition probabilities at different forces from eq.~(\ref{eq:occtot}). The results for realizations with a single optical lattice and a superlattice are shown in fig.~\ref{fig:singlelattice}. We clearly observe resonances of different orders as the external force is varied. The resonance position changes nonlinearly with the lattice depth since the band gap is generally a not strictly linear function of the lattice depth (in both cases of a single lattice and a superlattice). Additionally, the importance of the non-resonant background becomes more important for larger forces, i.e.\ for small $1/F$.

\begin{figure}
	\centering
	\resizebox{0.475\textwidth}{!}{\includegraphics{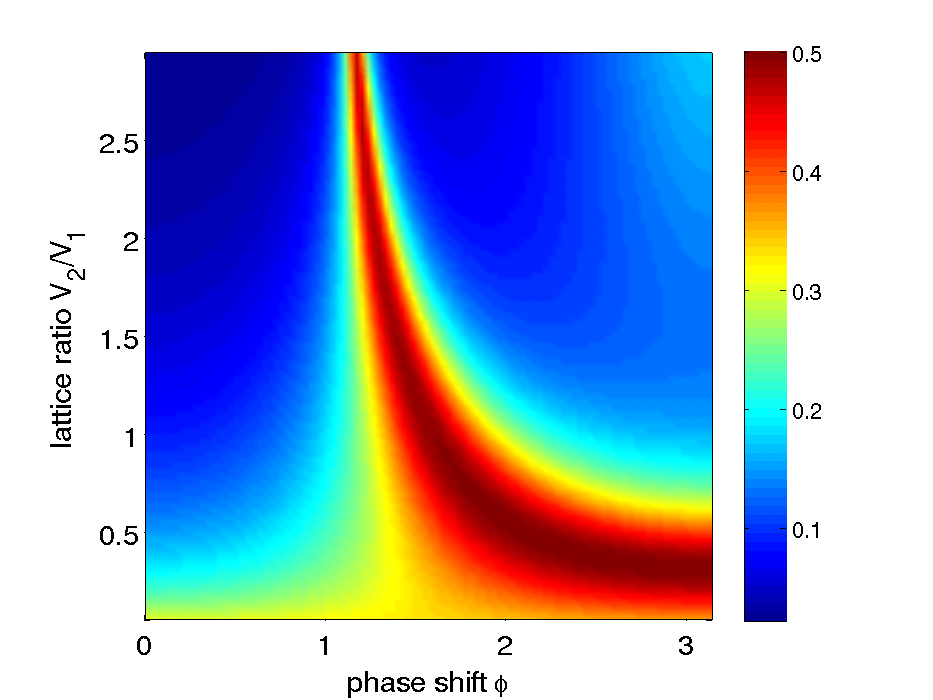}}
	\caption{(Color online) Driving interferometry, i.e. average occupation as function of the upper band as a function for a realization with a superlattice $V(x) = V_1\cos(x)+V_2\cos(2x+\phi)$. Shown is the average of the occupation of the upper band according to eq.~(\ref{eq:occtot}) as a function of the accessible experimental parameters: ratio of lattice depths $V_2/V_1$ and phase between the lattices $\phi$ for fixed external force $F=3$ and $V_1 =2$. We observe a clear and broad resonance from phase interference. The nonlinear shape arises from the nonlinear dependence of the various system parameters on $V_2/V_1$ and $\phi$.}
	\label{fig:superlattice}       
\end{figure}

As mentioned already, the three system parameters $J$, $\Delta$, and $C_0$ all depend on the lattice depth $V_0$ (or $V_2/V_1$) and cannot be varied independently. However, the situation is slightly advantageous for superlattices, since there are two experimental parameters, the ratio of the lattice depths $V_2/V_1$ and the relative phase between the lattices $\phi$, that can be altered. We therefore computed the Wannier functions and the system parameters for many different combinations of these two parameters and show the resulting transition probability as a contour plot in fig.~\ref{fig:superlattice} for fixed external force $F=3$.
We observe a clear and broad resonance as a result of St\"uckelberg interference when different parts of the wave function evolve in the different bands. The complicated shape of the resonance is again a result of the nonlinear dependence of the system parameters on $V_2/V_1$ and $\phi$. Fig.~\ref{fig:superlattice} is an explicit prediction for transition probabilities from multiple phase interference that should be observable with current experimental methods as in~\cite{salger}.

\section{Summary}
We have made the connection between St\"uckelberg interferometry and recent experiments with ultra-cold atoms. We showed explicitly how to obtain the Landau--Zener--St\"uckelberg Hamiltonian with cold atoms in accelerated optical lattices. More specifically, we applied the Magnus expansion to obtain analytical expressions capturing various aspects of the complicated interband dynamics. The transition probabilities for different experimental realizations with atomic quantum gases have been computed and should be experimentally accessible. We thus hope to have clarified some of the background of ongoing experiments and to stimulate further research using the high control in state-of-the-art implementations.

\begin{acknowledgements}
This work was supported by the DFG FOR760 and the Klaus Tschira Foundation. SW is especially grateful to the Hengstberger Foundation for the Klaus-Georg and Sigrid Hengstberger Prize, and acknowledges furthermore support from the Helmholtz Alliance Program of the Helmholtz Association (contract HA-216 ``Extremes of Density and Temperature: Cosmic Matter in the Laboratory''), and within the framework of the Excellence Initiative by the German Research Foundation (DFG) through the Heidelberg Graduate School of Fundamental Phys\-ics (grant number GSC 129/1), the Frontier Innovation Fonds and the Global Networks Mobility Measures.
\end{acknowledgements}

\bibliography{biblio}

\begin{thebibliography}{23}
\expandafter\ifx\csname natexlab\endcsname\relax\def\natexlab#1{#1}\fi
\expandafter\ifx\csname bibnamefont\endcsname\relax
  \def\bibnamefont#1{#1}\fi
\expandafter\ifx\csname bibfnamefont\endcsname\relax
  \def\bibfnamefont#1{#1}\fi
\expandafter\ifx\csname citenamefont\endcsname\relax
  \def\citenamefont#1{#1}\fi
\expandafter\ifx\csname url\endcsname\relax
  \def\url#1{\texttt{#1}}\fi
\expandafter\ifx\csname urlprefix\endcsname\relax\def\urlprefix{URL }\fi
\providecommand{\bibinfo}[2]{#2}
\providecommand{\eprint}[2][]{\url{#2}}

\bibitem[{\citenamefont{Grifoni and H\"anggi}(1998)}]{grifoni}
\bibinfo{author}{\bibfnamefont{M.}~\bibnamefont{Grifoni}} \bibnamefont{and}
  \bibinfo{author}{\bibfnamefont{P.}~\bibnamefont{H\"anggi}},
  \bibinfo{journal}{Physics Reports} \textbf{\bibinfo{volume}{304}},
  \bibinfo{pages}{229 } (\bibinfo{year}{1998}).

\bibitem[{\citenamefont{Oliver et~al.}(2005)\citenamefont{Oliver, Yu, Lee,
  Berggren, Levitov, and Orlando}}]{squbit}
\bibinfo{author}{\bibfnamefont{W.~D.} \bibnamefont{Oliver}},
  \bibinfo{author}{\bibfnamefont{Y.}~\bibnamefont{Yu}},
  \bibinfo{author}{\bibfnamefont{J.~C.} \bibnamefont{Lee}},
  \bibinfo{author}{\bibfnamefont{K.~K.} \bibnamefont{Berggren}},
  \bibinfo{author}{\bibfnamefont{L.~S.} \bibnamefont{Levitov}},
  \bibnamefont{and} \bibinfo{author}{\bibfnamefont{T.~P.}
  \bibnamefont{Orlando}}, \bibinfo{journal}{Science}
  \textbf{\bibinfo{volume}{310}}, \bibinfo{pages}{1653} (\bibinfo{year}{2005}).

\bibitem[{\citenamefont{Longhi}(2010)}]{Longhi}
\bibinfo{author}{\bibfnamefont{S.}~\bibnamefont{Longhi}},
  \bibinfo{journal}{Phys. Rev. A} \textbf{\bibinfo{volume}{81}},
  \bibinfo{pages}{022118} (\bibinfo{year}{2010}).

\bibitem[{\citenamefont{Shevchenko et~al.}(2010)\citenamefont{Shevchenko,
  Ashhab, and Nori}}]{shevchenko}
\bibinfo{author}{\bibfnamefont{S.}~\bibnamefont{Shevchenko}},
  \bibinfo{author}{\bibfnamefont{S.}~\bibnamefont{Ashhab}}, \bibnamefont{and}
  \bibinfo{author}{\bibfnamefont{F.}~\bibnamefont{Nori}},
  \bibinfo{journal}{Physics Reports} \textbf{\bibinfo{volume}{492}},
  \bibinfo{pages}{1 } (\bibinfo{year}{2010}).

\bibitem[{\citenamefont{Zenesini et~al.}(2010)\citenamefont{Zenesini, Ciampini,
  Morsch, and Arimondo}}]{arimondo}
\bibinfo{author}{\bibfnamefont{A.}~\bibnamefont{Zenesini}},
  \bibinfo{author}{\bibfnamefont{D.}~\bibnamefont{Ciampini}},
  \bibinfo{author}{\bibfnamefont{O.}~\bibnamefont{Morsch}}, \bibnamefont{and}
  \bibinfo{author}{\bibfnamefont{E.}~\bibnamefont{Arimondo}},
  \bibinfo{journal}{arXiv:1010.2431v1}  (\bibinfo{year}{2010}).

\bibitem[{\citenamefont{Kling et~al.}(2010)\citenamefont{Kling, Salger,
  Grossert, and Weitz}}]{salger}
\bibinfo{author}{\bibfnamefont{S.}~\bibnamefont{Kling}},
  \bibinfo{author}{\bibfnamefont{T.}~\bibnamefont{Salger}},
  \bibinfo{author}{\bibfnamefont{C.}~\bibnamefont{Grossert}}, \bibnamefont{and}
  \bibinfo{author}{\bibfnamefont{M.}~\bibnamefont{Weitz}},
  \bibinfo{journal}{Phys. Rev. Lett.} \textbf{\bibinfo{volume}{105}},
  \bibinfo{pages}{215301} (\bibinfo{year}{2010}).

\bibitem[{\citenamefont{Morsch and Oberthaler}(2006)}]{OberthalerReview}
\bibinfo{author}{\bibfnamefont{O.}~\bibnamefont{Morsch}} \bibnamefont{and}
  \bibinfo{author}{\bibfnamefont{M.}~\bibnamefont{Oberthaler}},
  \bibinfo{journal}{Rev. Mod. Phys.} \textbf{\bibinfo{volume}{78}},
  \bibinfo{eid}{179} (\bibinfo{year}{2006}).

\bibitem[{\citenamefont{Bloch et~al.}(2008)\citenamefont{Bloch, Dalibard, and
  Zwerger}}]{BlochZwergerReview}
\bibinfo{author}{\bibfnamefont{I.}~\bibnamefont{Bloch}},
  \bibinfo{author}{\bibfnamefont{J.}~\bibnamefont{Dalibard}}, \bibnamefont{and}
  \bibinfo{author}{\bibfnamefont{W.}~\bibnamefont{Zwerger}},
  \bibinfo{journal}{Rev. Mod. Phys.} \textbf{\bibinfo{volume}{80}},
  \bibinfo{pages}{885} (\bibinfo{year}{2008}).

\bibitem[{\citenamefont{Tomadin et~al.}(2008)\citenamefont{Tomadin, Mannella,
  and Wimberger}}]{TMW1}
\bibinfo{author}{\bibfnamefont{A.}~\bibnamefont{Tomadin}},
  \bibinfo{author}{\bibfnamefont{R.}~\bibnamefont{Mannella}}, \bibnamefont{and}
  \bibinfo{author}{\bibfnamefont{S.}~\bibnamefont{Wimberger}},
  \bibinfo{journal}{Phys. Rev. A} \textbf{\bibinfo{volume}{77}},
  \bibinfo{eid}{013606} (\bibinfo{year}{2008}).

\bibitem[{\citenamefont{Tomadin}(2006)}]{andrea}
\bibinfo{author}{\bibfnamefont{A.}~\bibnamefont{Tomadin}}, Master's thesis,
  \bibinfo{school}{Universit\`{a} di Pisa} (\bibinfo{year}{2006}).

\bibitem[{\citenamefont{Pl\"{o}tz}(2010)}]{diss}
\bibinfo{author}{\bibfnamefont{P.}~\bibnamefont{Pl\"{o}tz}}, Ph.D. thesis,
  \bibinfo{school}{Universit\"{a}t Heidelberg} (\bibinfo{year}{2010}),
  \urlprefix\url{http://archiv.ub.uni-heidelberg.de/volltextserver/volltexte/2%
010/11123/}.

\bibitem[{\citenamefont{Parra-Murillo et~al.}()\citenamefont{Parra-Murillo,
  Madro{\~n}ero, and Wimberger}}]{carlos}
\bibinfo{author}{\bibfnamefont{C.}~\bibnamefont{Parra-Murillo}},
  \bibinfo{author}{\bibfnamefont{J.}~\bibnamefont{Madro{\~n}ero}},
  \bibnamefont{and}
  \bibinfo{author}{\bibfnamefont{S.}~\bibnamefont{Wimberger}},
  \bibinfo{note}{in preparation.}

\bibitem[{\citenamefont{Pl{\"o}tz et~al.}(2010)\citenamefont{Pl{\"o}tz,
  Madro{\~n}ero, and Wimberger}}]{collapse}
\bibinfo{author}{\bibfnamefont{P.}~\bibnamefont{Pl{\"o}tz}},
  \bibinfo{author}{\bibfnamefont{J.}~\bibnamefont{Madro{\~n}ero}},
  \bibnamefont{and}
  \bibinfo{author}{\bibfnamefont{S.}~\bibnamefont{Wimberger}},
  \bibinfo{journal}{J.\ Phys.\ B} \textbf{\bibinfo{volume}{43}},
  \bibinfo{pages}{081001} (\bibinfo{year}{2010}).

\bibitem[{\citenamefont{Pl{\"o}tz et~al.}()\citenamefont{Pl{\"o}tz, Schlagheck,
  and Wimberger}}]{spinmodel}
\bibinfo{author}{\bibfnamefont{P.}~\bibnamefont{Pl{\"o}tz}},
  \bibinfo{author}{\bibfnamefont{P.}~\bibnamefont{Schlagheck}},
  \bibnamefont{and}
  \bibinfo{author}{\bibfnamefont{S.}~\bibnamefont{Wimberger}},
  \emph{\bibinfo{title}{Complex dynamics of a two-band bose--hubbard model}},
  \bibinfo{note}{in preparation}.

\bibitem[{\citenamefont{Kolovsky and Buchleitner}(2003)}]{FB}
\bibinfo{author}{\bibfnamefont{A.~R.} \bibnamefont{Kolovsky}} \bibnamefont{and}
  \bibinfo{author}{\bibfnamefont{A.}~\bibnamefont{Buchleitner}},
  \bibinfo{journal}{Phys. Rev. E} \textbf{\bibinfo{volume}{68}},
  \bibinfo{pages}{056213} (\bibinfo{year}{2003}).

\bibitem[{\citenamefont{Zhao et~al.}(1996)\citenamefont{Zhao, Georgakis, and
  Niu}}]{Zhao}
\bibinfo{author}{\bibfnamefont{X.-G.} \bibnamefont{Zhao}},
  \bibinfo{author}{\bibfnamefont{G.~A.} \bibnamefont{Georgakis}},
  \bibnamefont{and} \bibinfo{author}{\bibfnamefont{Q.}~\bibnamefont{Niu}},
  \bibinfo{journal}{Phys. Rev. B} \textbf{\bibinfo{volume}{54}},
  \bibinfo{pages}{R5235} (\bibinfo{year}{1996}).

\bibitem[{\citenamefont{Pl{\"o}tz}(2010)}]{noninter}
\bibinfo{author}{\bibfnamefont{P.}~\bibnamefont{Pl{\"o}tz}},
  \bibinfo{journal}{Journal of Siberian Federal University: Mathematics \&
  Physics} \textbf{\bibinfo{volume}{3}}, \bibinfo{pages}{381}
  (\bibinfo{year}{2010}).

\bibitem[{\citenamefont{Blanes et~al.}(2009)\citenamefont{Blanes, Casas, Oteo,
  and Ros}}]{MagnusReview}
\bibinfo{author}{\bibfnamefont{S.}~\bibnamefont{Blanes}},
  \bibinfo{author}{\bibfnamefont{F.}~\bibnamefont{Casas}},
  \bibinfo{author}{\bibfnamefont{J.}~\bibnamefont{Oteo}}, \bibnamefont{and}
  \bibinfo{author}{\bibfnamefont{J.}~\bibnamefont{Ros}},
  \bibinfo{journal}{Physics Reports} \textbf{\bibinfo{volume}{470}},
  \bibinfo{pages}{151 } (\bibinfo{year}{2009}).

\bibitem[{\citenamefont{Rojo}(2010)}]{rojo}
\bibinfo{author}{\bibfnamefont{A.~G.} \bibnamefont{Rojo}},
  \bibinfo{journal}{arXiv:1004.2914v1}  (\bibinfo{year}{2010}).

\bibitem[{\citenamefont{Pechukas and Light}(1966)}]{MagnusPechukas}
\bibinfo{author}{\bibfnamefont{P.}~\bibnamefont{Pechukas}} \bibnamefont{and}
  \bibinfo{author}{\bibfnamefont{J.~C.} \bibnamefont{Light}},
  \bibinfo{journal}{The Journal of Chemical Physics}
  \textbf{\bibinfo{volume}{44}}, \bibinfo{pages}{3897} (\bibinfo{year}{1966}).

\bibitem[{\citenamefont{Gl{\"u}ck et~al.}(2002)\citenamefont{Gl{\"u}ck,
  Kolovsky, and Korsch}}]{WSreview}
\bibinfo{author}{\bibfnamefont{M.}~\bibnamefont{Gl{\"u}ck}},
  \bibinfo{author}{\bibfnamefont{A.~R.} \bibnamefont{Kolovsky}},
  \bibnamefont{and} \bibinfo{author}{\bibfnamefont{H.~J.}
  \bibnamefont{Korsch}}, \bibinfo{journal}{Physics Reports}
  \textbf{\bibinfo{volume}{366}}, \bibinfo{pages}{103} (\bibinfo{year}{2002}).

\bibitem[{\citenamefont{Shirley}(1963)}]{ShirleyThesis}
\bibinfo{author}{\bibfnamefont{J.~H.} \bibnamefont{Shirley}}, Ph.D. thesis,
  \bibinfo{school}{California Institute of Technology} (\bibinfo{year}{1963}).

\bibitem[{\citenamefont{Hausinger and Grifoni}(2010)}]{twolevelsystem}
\bibinfo{author}{\bibfnamefont{J.}~\bibnamefont{Hausinger}} \bibnamefont{and}
  \bibinfo{author}{\bibfnamefont{M.}~\bibnamefont{Grifoni}},
  \bibinfo{journal}{Phys. Rev. A} \textbf{\bibinfo{volume}{81}},
  \bibinfo{pages}{022117} (\bibinfo{year}{2010}).

\end{thebibliography}

\end{document}